\documentclass[twocolumn,showpacs,amsmath,amssymb]{revtex4}
\usepackage{graphicx}

\newcommand{\bg}{\mbox{\boldmath $\gamma$}}

\begin{document}

\title{Embedding Quantum Information into Classical Spacetime: Information Geometrical Perspectives on anti-de Sitter space / conformal field theory Correspondence}
\author{Hiroaki Matsueda\footnote{matsueda@sendai-nct.ac.jp}}
\affiliation{Sendai National College of Technology, Sendai 989-3128, Japan}
\date{\today}
\begin{abstract}
An information-geometrical interpretation of AdS${}_{3}$/CFT${}_{2}$ correspondence is given. In particular, we consider an inverse problem in which the classical spacetime metric is given in advance and then we find what is the proper quantum information that is well stored into the spacetime. We see that the Fisher metric plays a central role on this problem. Actually, if we start with the two-dimensional hyperbolic space, a constant-time surface in AdS${}_{3}$, the resulting singular value spectrum of the quantum state shows power law for the correlation length with conformal dimension proportional to the curvature radius in the gravity side. Furthermore, the entanglement entropy data embedded into the hyperbolic space agree well with the Ryu-Takayanagi formula. These results show that the relevance of the AdS/CFT correspondence can be represented by the information-gemetrical approach based on the Fisher metric.
\end{abstract}
\maketitle

\section{Introduction}

Transforming a complicated quantum system into classical one provides a lot of useful insights into wide area of theoretical physics. A prototypical example in string-theory community is the anti-de Sitter space / conformal field theory (AdS/CFT) correspondence~\cite{Maldacena}. In statistical physics, the multiscale entanglement renormalization ansatz (MERA) is an efficient variational formulation for quantum critical models~\cite{Vidal}, and it is an interesting topic to examine the similarity between MERA and AdS/CFT~\cite{Swingle}. Even apart from these theories, the Suzuki-Trotter breakup has been a very important quantum-classical correspondence for a long time~\cite{Suzuki}, and recently its close connection to AdS/CFT and MERA has been discussed~\cite{Matsueda}. It is thus attracting attention to know common mathematical properties behind them.

Some aspects of those correspondences are deeply intertwined with data storage. In information-theoretical terminology, the question seems to find a proper way of data storage or to find a transformation method of data format so that the information entropy is conserved. It is clear in terms of AdS/CFT that the conservation of the total amount of the data after the quantum-classical transformation is represented by the famous Ryu-Takayanagi's holographic formula for the entanglement entropy~\cite{Ryu}. Therefore, we call quantum information or reduced density matrix spectra as 'quantum data' to be stored, while we call the classical spacetime as 'memory space (spacetime)' (in general the memory would be time-dependent).

In this respect, it is very efficient to examine geometrical structure of the memory space by based on the Fisher information~\cite{Amari}. As we will introduce the definition, the Fisher information classically measures a difference between two quantum states by comparing between their sets of the reduced-density-matrix spectra. By naturally following from the underlying properties of the density-matrix spectra, the memory space should be a manifold with an appropriate metric in terms of Riemaniann geometry. Thus, it is not only able to introduce the metric in the memory space, but also connect the original quantum state with the classical memory space in a quite natural way. That is related to the idea of AdS/CFT. Actually, the close connection between MERA and AdS/CFT has been recently examined by using the Bures metric, a kind of Fisher information~\cite{Nozaki,Mollabashi}. Thus if quantum data are first prepared, it is possible to examine what kind of metric emerges from the quantum data.

Motivated by such consideration, here we examine an inverse problem in which the metric of the memory space is given {\it in advance} and then we find what is the original quantum information that is properly stored into the memory space. The approach starting from some quantum state like~\cite{Nozaki,Mollabashi} is quite straightforwad, but at the same time the inverse approach is also necessary for more conclete understanding of this type of problem. Our objective is to know how strongly the geometry, a kind of 'receptacle', restricts the format of acceptable data. For simplicity, we focus on a time-constant surface of AdS${}_{3}$/CFT${}_{2}$, that is correspondence between the CFT data in equillibrium and two-dimensional (2D) hyperbolic space. We will find that the geometry can determine the entanglement structure of the original quantum data to be stored. The result shows that the Fisher-information approach is powerful to examine the AdS/CFT correspondence.

The paper is organized as follows. In Sec.~II, general set up of our quantum state and its complex coordinate representation are given. In Sec.~III, we define the Fisher metric, and develop complex function analysis to dermine the entanglement spectrum so that the metric becomes hyperbolic. Finally, our results are summarized in Sec.~VI.

\section{General Setup in Quantum Side}

\subsection{Entanglement Spectrum and Entropy}

We start with any quantum state in spatially one dimension represented by the Schmidt decomposition or the singular value decomposition (SVD)
\begin{eqnarray}
\left|\psi\right> &=& \sum_{i,j}U_{n}(i)\sqrt{\lambda_{n}}V_{n}(j)\left|i\right>_{A}\otimes\left|j\right>_{\bar{A}} \nonumber \\
&=& \sum_{n}\sqrt{\lambda_{n}}\left|n\right>_{A}\otimes\left|n\right>_{\bar{A}},
\end{eqnarray}
where the whole system with finite size $L$ is devided into a subsystem $A$ and its complement $\bar{A}$. We particulary focus on an equilibrium state ($t=0$). Then, the SVD spectrum $\lambda_{n}$ is a function of the correlation length $\xi$ and the boundary position $\eta$ between $A$ and $\bar{A}$ as
\begin{eqnarray}
\lambda_{n}=\lambda_{n}(\xi,\eta)=\lambda_{n}(x),
\end{eqnarray}
where we take $x=(x^{0},x^{1})=(\xi,\eta)$. It is well-known from DMRG that each Schmidt basis represent a particular length-scale physics~\cite{White}. When we restrict the sum of the decomposition up to $\chi$, then the state is away from criticality and has a finite correlation length $\xi$. Thus, $\xi$ appears quite naturally. Now, there is one boundary point for a specific choice of $A$. However, we can freely change the length of $A$. Thus, the boundary coordinate $\eta$ runs over whole 1D system with size $L$, and we can regard $\eta$ as usual spatial coordinate in the whole system. We should be careful for a fact that we are going to transform the amount of entanglement that each Schmidt decomposition form with a specific $A$ has into a classical representation. We introduce $A$ and $\bar{A}$ in order to just define the entanglement properties inside of $A+\bar{A}$. More precisely speaking, the devision of the system at $\eta=L/2$ has maximal entanglement, while taking $\eta\rightarrow 0$ or $\eta\rightarrow L$ reduces the entanglement that can be represented by the Schmidt decomposition. Such a kind of entanglement data will be embedded into the gravity side. Here, it is worth mentioning that $\xi>\eta$ near the critical point. At the critical point, we can take $\xi\rightarrow L$, since the size of the whole system is restricted to be $L$. Thus, the $\eta$ dependence on the SVD spectrum and the entanglement entropy are not usually important in the critical point, but we should formally consider the dependence for general quantum states. For later convenience, we normalize $\lambda_{n}$ as
\begin{eqnarray}
\sum_{n}\lambda_{n}(x)=1.
\end{eqnarray}
We define the entanglement spectrum $\gamma_{n}(x)$ and the entanglement entropy $S(x)$ as
\begin{eqnarray}
\gamma_{n}(x) &=& -\log\lambda_{n}(x) , \\
S(x) &=& -\sum_{n}\lambda_{n}(x)\log\lambda_{n}(x)=\left<\bg\right> , \label{EE}
\end{eqnarray}
where we have defined the avarage of a quantity $O_{n}(x)$ as
\begin{eqnarray}
\left<O\right>=\sum_{n}\lambda_{n}(x)O_{n}(x).
\end{eqnarray}

\subsection{Complex Coordinate}

A key procedure toward the next section is to represent $\gamma_{n}(x)$ as complex coordinate. For this purpose, we introduce the coordinate as
\begin{eqnarray}
w &=& x^{1} + ix^{0} = \eta + i\xi , \\
\bar{w} &=& x^{1} - ix^{0} = \eta - i\xi ,
\end{eqnarray}
and thus we denote $\gamma_{n}(\xi,\eta)$ as $\gamma_{n}(w,\bar{w})$. They cover the upper half of the whole complex plane due to the condition $\xi>0$. Hereafter, we treat complex functions that can be analytically continued to the other half of the whole complex plane except for the origin if necessary. 

Since $\gamma_{n}(x)$ is a real-valued two-parameter function, the complex coordinate representation yields that $\gamma_{n}(x)$ is sum of holomorphic and anti-holomorphic function. Let $f_{n}(w)$ be a holomorphic function with index $n$. We consider a situation that the first derivative of $f_{n}(w)$ has the following Laurent series expansion
\begin{eqnarray}
\partial_{w}f_{n}(w) = \sum_{j}A_{n}^{(j)}w^{j},
\end{eqnarray}
Then, $f_{n}(w)$ is as follows
\begin{eqnarray}
f_{n}(w) = \cdots + a_{n} + A_{n}^{(-1)}\log w - \frac{A_{n}^{(-2)}}{w} + \cdots .
\end{eqnarray}
with a constant $a_{n}$. For the case that $\xi>\eta$, the logarithmic function is expanded to be
\begin{eqnarray}
\log w = \log\left(\eta+i\xi\right) = \log\left(i\xi\right)+\frac{\eta}{i\xi}+\cdots. \label{logw}
\end{eqnarray}
Then, we obtain
\begin{eqnarray}
f_{n}(w) &=& \cdots + \left(a_{n}+A_{n}^{(-1)}\frac{\nu\pi i}{2}\right) \nonumber \\
&& + A_{n}^{(-1)}\log\xi + A_{n}^{(-1)}\frac{\eta}{i\xi} + \cdots ,
\end{eqnarray}
where $\nu$ is an odd integer. Next we introduce the anti-holomorphic part of $f_{n}(w)$ as
\begin{eqnarray}
\bar{f}_{n}(\bar{w}) &=& \cdots + \left(\bar{a}_{n}-\bar{A}_{n}^{(-1)}\frac{\nu\pi i}{2}\right) \nonumber \\
&& + \bar{A}_{n}^{(-1)}\log\xi - \bar{A}_{n}^{(-1)}\frac{\eta}{i\xi} + \cdots .
\end{eqnarray}
The sum of $f_{n}(w)$ and $\bar{f}_{n}(\bar{w})$ yields
\begin{eqnarray}
\gamma_{n}(w,\bar{w}) &=& f_{n}(w)+\bar{f}_{n}(\bar{w}) \nonumber \\
&=& \cdots + \left(2\Re a_{n} - \nu\pi\Im A_{n}^{(-1)}\right) \nonumber \\
&& + 2\Re A_{n}^{(-1)}\log\xi + 2\Im A_{n}^{(-1)}\frac{\eta}{\xi} + \cdots . \label{fwfw}
\end{eqnarray}

This form is quite suggestive for geometrical meaning of the entanglement spectrum, when the original quantum system is near a quantum critical point. Remember the Calabrese-Cardy formula for the entanglement entropy represented as
\begin{eqnarray}
S=\frac{c}{6}\log\xi,
\end{eqnarray}
where $c$ is the central charge of the quantum model (this is correct for a case with one boundary point)~\cite{Holzhey,Calabrese}. Since this entropy is obtained as a result of avaraging procedure for the entanglement spectrum, we can expect that the logarithmic terms in $f_{n}(w)$ and $\bar{f}_{n}(\bar{w})$ produce the entropy formula. Then, we identify $A_{n}^{(-1)}$ as
\begin{eqnarray}
\Re A_{n}^{(-1)} = \frac{c}{12} \;\; , \;\; \Im A_{n}^{(-1)}=0.
\end{eqnarray}
Therefore, the magnitude of the real part of $A_{n}^{(-1)}$ characterizes the central extention of the Virasoro algebra.

The reader might think that the above procedure is too artificial. We just pick up the logarithmic term from the Laurent series. However, as we will see later, the function $\partial_{w}\gamma_{n}(w,\bar{w})$ plays a central role on the construction of the metric in dual gravity side.

\section{Geometric Structure of the Memory Space}

\subsection{Relative Entanglement Entropy and Fisher Metric}

The complex coordinate representation is deeply intertwined with the construction of the spacetime metric relevant for data embedding of quantum entanglement. For this purpose, we take the second derivative of the entanglement entropy $S(x)$, and then obtain
\begin{eqnarray}
-\partial_{\mu}\partial_{\nu}S(x) &=& \sum_{n}\frac{1}{\lambda_{n}(x)}\partial_{\mu}\lambda_{n}(x)\partial_{\nu}\lambda_{n}(x) \nonumber \\
&=& \sum_{n}\lambda_{n}(x)\partial_{\mu}\gamma_{n}(x)\partial_{\nu}\gamma_{n}(x) \nonumber \\
&=& \left<\partial_{\mu}\gamma\partial_{\nu}\gamma\right>. \label{mnS}
\end{eqnarray}
The right hand side of Eq.~(\ref{mnS}) is nothing but the Fisher metric in terms of information geometry~\cite{Amari}. Actually, we calculate the infinitesimal change of the entropy, the so-called Kullback-Leibler measure, as
\begin{eqnarray}
D(x) &=& \sum_{n}\lambda_{n}(x)\left(\gamma_{n}(x)-\gamma_{n}(x+dx)\right) \nonumber \\
&=& \frac{1}{2}\left<\partial_{\mu}\gamma\partial_{\nu}\gamma\right>dx^{\mu}dx^{\nu}, \label{KL}
\end{eqnarray}
and we find that
\begin{eqnarray}
g_{\mu\nu}(x)=k^{2}\left<\partial_{\mu}\gamma\partial_{\nu}\gamma\right>=-k^{2}\partial_{\mu}\partial_{\nu}S(x),
\end{eqnarray}
where $k$ is the unit of length that will be later determined.

\subsection{Data Structure Efficiently Stored into Hyperbolic Space}

As already mentioned, the metric is a correlation function between $\partial_{\mu}\gamma_{n}$ and $\partial_{\nu}\gamma_{n}$. Thus, the complex coordinate becomes powerful to properly represent geometric properties of the memory space.

Here, our goal is to find the quantum data structure that can be efficiently stored into the hyperbolic space. That is, when we assume the following form
\begin{eqnarray}
ds^{2}=g_{\mu\nu}dx^{\mu}dx^{\nu} = l^{2}\frac{d\xi^{2}+d\eta^{2}}{\xi^{2}} = l^{2}\frac{dwd\bar{w}}{({\rm Im}w)^{2}},
\end{eqnarray}
we would like to examine whether the CFT nature automatically appears in the expression of $\bg$. In this sense, our problem is a kind of the inverse problem. For $\xi>\eta$, we can approximately treat this as
\begin{eqnarray}
ds^{2}\simeq l^{2}\frac{dwd\bar{w}}{|w|^{2}}. \label{complexmetric}
\end{eqnarray}
With use of complex coordinate for $\bg$, the Fisher metric is represented as
\begin{eqnarray}
g_{\mu\nu}dx^{\mu}dx^{\nu} &=& k^{2}\left<\frac{\partial\gamma}{\partial w}\frac{\partial\gamma}{\partial w}\right>dw^{2} + k^{2}\left<\frac{\partial\gamma}{\partial\bar{w}}\frac{\partial\gamma}{\partial\bar{w}}\right>d\bar{w}^{2} \nonumber \\
&& + 2k^{2}\left<\frac{\partial\gamma}{\partial w}\frac{\partial\gamma}{\partial\bar{w}}\right>dwd\bar{w} .
\end{eqnarray}
Therefore, our problem is to solve the following equation for $\bg$
\begin{eqnarray}
l^{2}\frac{dwd\bar{w}}{|w|^{2}}  &=& k^{2}\left<\frac{\partial\gamma}{\partial w}\frac{\partial\gamma}{\partial w}\right>dw^{2} + k^{2}\left<\frac{\partial\gamma}{\partial\bar{w}}\frac{\partial\gamma}{\partial\bar{w}}\right>d\bar{w}^{2} \nonumber \\
&& + 2k^{2}\left<\frac{\partial\gamma}{\partial w}\frac{\partial\gamma}{\partial\bar{w}}\right>dwd\bar{w} . \label{problem}
\end{eqnarray}

Let us solve Eq.~(\ref{problem}). It is almost clear that the solution comes from the $A_{n}^{(-1)}$ term in the Laurent series expansion of $\bg$. Thus, we take the following form
\begin{eqnarray}
\gamma_{n}(w,\bar{w}) &=& g_{n} + h_{n}\ln w + \bar{h}_{n}\ln\bar{w}, \label{gamma3}
\end{eqnarray}
where $\gamma_{n}(w,\bar{w})$ should be real, and thus is decomposed into sum of holomorphic and anti-holomorphic parts, $h_{n}$ is a complex variable, and $g_{n}$ is real. The SVD spectrum is then represented as
\begin{eqnarray}
\lambda_{n}(w,\bar{w}) = e^{-\gamma_{n}(w,\bar{w})} = e^{-g_{n}}w^{-h_{n}}\bar{w}^{-\bar{h}_{n}} .
\end{eqnarray}
When we introduce a conformal transformation given by $w\rightarrow f(w)$ and $\bar{w}\rightarrow f(\bar{w})$, the SVD spectrum $\lambda_{n}(w,\bar{w})$ is transformed into the following form
\begin{eqnarray}
\lambda_{n}(f(w),f(\bar{w})) &=& e^{-g_{n}}f(w)^{-h_{n}}f(\bar{w})^{-\bar{h}_{n}} \nonumber \\
&=& \left(\frac{f(w)}{w}\right)^{\!\! -h_{n}}\!\!\left(\frac{f(\bar{w})}{\bar{w}}\right)^{\!\! -\bar{h}_{n}}\!\!\lambda_{n}(w,\bar{w}) . \label{trans} \nonumber \\
\end{eqnarray}
Thus, $\lambda_{n}$ behaves as a primary-like field, if the mapping $f$ only changes the scale $f(w)=Aw$. In this case, we have $\partial f(w)/\partial w = f(w)/w$. The parameter
\begin{eqnarray}
\Delta_{n}=h_{n} + \bar{h}_{n},
\end{eqnarray}
or the pair $(h_{n},\bar{h}_{n})$ is the conformal (scaling) dimension.

Next we determine $h_{n}$ and $g_{n}$. We represent $h_{n}$ with use of two real variables $\alpha_{n}$ and $\beta_{n}$ as
\begin{eqnarray}
h_{n} = \alpha_{n} + i\beta_{n} , \label{h}
\end{eqnarray}
and we take
\begin{eqnarray}
|\alpha_{n}|^{2}=|\beta_{n}|^{2} , \label{as1}
\end{eqnarray}
in order to remove some terms in the later processes. By using the above notations, the each component of the line element is represented as
\begin{eqnarray}
\left<\frac{\partial\gamma}{\partial w}\frac{\partial\gamma}{\partial w}\right> &=& \frac{\left<h^{2}\right>}{w^{2}} = 2i\frac{\left<\alpha\beta\right>}{w^{2}} , \\
\left<\frac{\partial\gamma}{\partial\bar{w}}\frac{\partial\gamma}{\partial\bar{w}}\right> &=& \frac{\left<\bar{h}^{2}\right>}{\bar{w}^{2}} = -2i\frac{\left<\alpha\beta\right>}{\bar{w}^{2}} ,
\end{eqnarray}
and
\begin{eqnarray}
\left<\frac{\partial\gamma}{\partial w}\frac{\partial\gamma}{\partial\bar{w}}\right> = \frac{\left<h\bar{h}\right>}{|w|^{2}} = \frac{\left<\alpha^{2}\right>+\left<\beta^{2}\right>}{|w|^{2}} = 2\frac{\left<\alpha^{2}\right>}{|w|^{2}} .
\end{eqnarray}
Then, we find
\begin{eqnarray}
ds^{2} = 2ik^{2}\left<\alpha\beta\right>\left(\frac{dw^{2}}{w^{2}}-\frac{d\bar{w}^{2}}{\bar{w}^{2}}\right) + 4k^{2}\left<\alpha^{2}\right>\frac{dwd\bar{w}}{|w|^{2}} .
\end{eqnarray}
Assuming the condition
\begin{eqnarray}
\left<\alpha\beta\right>=0, \label{as3}
\end{eqnarray}
we obtain
\begin{eqnarray}
ds^{2} = 4k^{2}\left<\alpha^{2}\right>\frac{dwd\bar{w}}{|w|^{2}} ,
\end{eqnarray}
and thus we know from Eq.~(\ref{complexmetric})
\begin{eqnarray}
4k^{2}\left<\alpha^{2}\right> = l^{2}. \label{as2}
\end{eqnarray}
By using Eq.~(\ref{logw}), $\gamma_{n}(w,\bar{w})$ is represented as
\begin{eqnarray}
\gamma_{n}(w,\bar{w}) &=& g_{n} - \beta_{n}\nu\pi + 2\alpha_{n}\ln\xi + 2\beta_{n}\frac{\eta}{\xi} \nonumber \\
&=& b_{n} + \Delta_{n}\ln\xi + a_{n}\frac{\eta}{\xi},
\end{eqnarray}
where
\begin{eqnarray}
\Delta_{n} &=& 2\alpha_{n} = \frac{l}{k}, \\
a_{n} &=& 2\beta_{n} , \\
b_{n} &=& g_{n}-\beta_{n}\nu\pi .
\end{eqnarray}
We also know that Eq.~(\ref{as3}) corresponds to $\left<a\right>=0$.

Finally, the SVD spectrum and the entanglement entropy are respectively given by
\begin{eqnarray}
\lambda_{n}(\xi,\eta) = e^{-\gamma_{n}(\xi,\eta)} = \frac{1}{\xi^{\Delta_{n}}}\exp\left(-b_{n}-a_{n}\frac{\eta}{\xi}\right) , \label{lambda}
\end{eqnarray}
and
\begin{eqnarray}
S = \left<\gamma\right> = \frac{l}{k}\log\xi + \left<g\right>. \label{entropy}
\end{eqnarray}
The first Eq.~(\ref{lambda}) means that $\lambda_{n}$ becomes very small toward the critical point $\xi\rightarrow\infty$ (or more precisely $\xi\rightarrow L$ in our setup), and then all of $\lambda_{n}$ with different $n$ play a role on the quantum state. Since the index $n$ distinguishes different length-scale physics, this means that all of them couple with each other. The second Eq.~(\ref{entropy}) is equivalent to the Ryu-Takayanagi formula, if we set the length scale $k$ as
\begin{eqnarray}
k=4G,
\end{eqnarray}
with the Newton constant $G$. All of them really match well with basic features of CFT. Therefore, what we can efficiently store in the memory space with hyperbolic coordinate are the CFT data. In other words, if we realize the AdS/CFT correspondence, the Fisher metric works well for the construction of dual gravity theory.

As for Eq.~(\ref{entropy}), we would like to remark that we are now looking at entanglement data just inside of the 1D system with finite size $L$. Thus, this entropy formula seems to observe the entanglement data inside of the minimal surface described by the Ryu-Takayanagi formula. In that sense, we are directly looking at the information itself rather than the minimal surface. This is slightly different from the original Ryu-Takayanagi formulation of the holographic entanglement entropy.

\section{Summary}

We have found that the appropriate quantum information should be the entanglement entropy data of CFT${}_{2}$ when we are going to efficiently embed them into the 2D hyperbolic space. In a viewpoint of information theory, we have first prepared the classical space as a receptacle or a memory, and then considered data structure stored properly into the space. The final results are consistent with the AdS/CFT correspondence. Actually, the resulting entanglement entropy in the gravity-side representation agrees well with the Ryu-Takayanagi formula, and the entanglement spectrum shows some properties required for quantum critical systems. The present complex analysis seems to be almost parallel to the standard CFT with one time and one space coordinates, except that our coordinate is defined by the space and radial coordinate. Only the difference is that our theory is formulated on a curved space and the appearance of quantum anomaly leads to the entanglement entropy formula. I stress that the Fisher information approach is quite powerful for deeply understanding AdS/CFT, and also think that the information-geometrical interpretation of the AdS/CFT correspondence is to find a way of matching between data structure and memory.

\end{document}